\documentclass[journal]{IEEEtran}

\usepackage[nolist]{acronym}
\usepackage{amsmath}
\usepackage[table,xcdraw]{xcolor}
\usepackage{balance}
\usepackage{tabularx}
\usepackage{subcaption}
\usepackage{multirow,array}
\newcolumntype{L}{>{\centering\arraybackslash}m{.32\columnwidth}}
\newcolumntype{M}{>{\centering\arraybackslash}m{.27\columnwidth}}

\usepackage{multicol}
 
\ifCLASSINFOpdf
\usepackage[pdftex]{graphicx}

\else
\fi
\renewenvironment{IEEEbiography}[1] 
{\IEEEbiographynophoto{#1}}
{\endIEEEbiographynophoto}

\begin{document}
\title{Towards scalable user-deployed ultra-dense networks: Blockchain-enabled small cells as a service}

\author{\IEEEauthorblockN{Emanuele Di Pascale, Hamed Ahmadi, Linda Doyle, and Irene Macaluso
\thanks{E. Di Pascale was with CONNECT centre for Future Networks and Communications, Trinity College Dublin, Ireland. He is now with Volta Networks.}
\thanks{H. Ahmadi is with the Department of Electronic Engineering, University of York, UK. }
\thanks{L. Doyle and I. Macaluso are with CONNECT centre for Future Networks and Communications, Trinity College Dublin, Ireland.}
\thanks{This material is based upon work supported by the Air force
Office of Scientific Research under award number FA9550-18-1-0214 and
supported by the Science Foundation Ireland under Grant No. 13/RC/2077.}
}

}


\maketitle
\bstctlcite{IEEEexample:BSTcontrol}
\begin{abstract}
Neutral Host Small Cell Providers (SCP) represent a key element of the 5G vision of ultra-dense mobile networks. However, current business models mostly focus on multi-year agreements for large venues, such as stadiums and hotel chains. These business agreements are regulated through binding Service Level Agreements (SLAs), which tend to be too cumbersome and costly for smaller scale SCPs. As a result, the neutral host model does not scale up to its full potential. In this paper, we propose a framework to enable the participation of small- to medium-sized players in the cellular market as providers offering network resources to Mobile Network Operators (MNOs). To this purpose, we review the current and emerging spectrum and technology opportunities that SCPs can use for neutral host deployments. We also propose the use of blockchain-enabled smart contracts as a simple and cost-efficient alternative to traditional SLAs for small-scale SCPs.  To demonstrate this, we describe a proof of concept implementation of an Ethereum-based smart contract platform for best-effort service between an SCP and an MNO. Our simulations on potential smart contract-based deployments in city centre Dublin show that the received signal strength in the considered area will increase by an average of $10$ percent.

\end{abstract}
\begin{IEEEkeywords}
	Smart contracts, small cell networks, neutral host, blockchain, service level agreement.
\end{IEEEkeywords}

\section{Introduction}
The neutral host model for small cell deployments is critical to meet the network ultra-densification required by fifth generation of mobile networks (5G), not only because of the cost reduction it affords, but also because it simplifies the logistics of the deployment. A neutral host Small Cell Provider (SCP) deploys and maintains a small cell network, and provides a common infrastructure that can be used by multiple Mobile Network Operators (MNOs) \cite{SF1}. 

The current business model for neutral host small cell deployments typically involves the SCP negotiating multi-year agreements with a large venue (e.g. class A buildings, stadiums, hotels) and providing a single point of contact to multiples MNOs. The SCP may maintain the ownership of the small cell network and lease access to the MNOs, or the MNO(s) may (partially) fund the deployment in exchange for a reduced usage fee. While the venues’ source of revenue is typically space rental, in some cases the venue management may act as an SCP itself. 

While there is a range of opportunities in terms of spectrum and technology for neutral host SCPs and  commercial solutions are already available \cite{ipAccess}\cite{MulteFire}, the scalability of this model is a key issue when one considers the legal and administrative burden of the agreements between SCPs and MNOs. User-deployed infrastructure as a means to enhance MNOs networks has been studied from a variety of angles: from technical aspects resulting from network densification with irregular deployments (see \cite{gotsis2016ultradense} and references therein) to options for incentivizing users’ participation \cite{iosifidis2014} to pricing mechanisms. 
The neutral host small cell model has also been investigated as part of 5G networks, once again focusing on technical issues, such as automated network planning \cite{munoz2018self} and analytics tools for resources optimization \cite{perez2018monitoring}.
However, to the best of our knowledge, mechanisms to address the scalability of establishing agreements with a large number of small-scale SCPs have not been investigated. This is a critical aspect when considering the number of medium and small commercial venues operating in the accommodation, food services, retail trade, and recreation sectors: according to data collected by the US Census Bureau and Eurostat, of the circa 7,400,000 establishments in US and EU, around 5,500,000 have less than 9 employees. The question then is how to enable owners of small cell infrastructure with small to medium footprints to participate in the cellular market as providers offering network resources to MNOs. Given the cost and logistics involved in deploying small cells, the ownership of the infrastructure is almost inevitably going to be highly fragmented. Rather than a hurdle, this can represent a huge opportunity. The main purpose of this work, therefore, is to suggest a possible framework to make the involvement of these small players affordable. We present blockchain-based smart contracts as means for enabling the small-scale SCPs to share their neutral host small cell base stations. 




The rest of the paper is organized as follows. In Section \ref{secII}, we review the current and emerging spectrum and technology opportunities that SCPs can use for neutral deployments. In Section \ref{Sec:SLA}, we introduce  Smart Contracts as possible enablers for reducing cost and difficulty of establishing long-term business agreements with a plethora of individual partners. As a proof of concept (PoC) of this idea, we describe our implementation of a best-effort smart contract platform over the Ethereum blockchain, in Section \ref{sec:PoC}. We then examine the implications in terms of received signal strength (RSS) in Section \ref{sec:implications}. Finally, in Section \ref{Sec:conc} we provide our concluding remarks and identify open issues that could be addressed in future works. 

\section{SPECTRUM AND TECHNOLOGY OPPORTUNITIES FOR NEUTRAL HOST SMALL CELL DEPLOYMENTS}\label{secII}

There are several options with respect to spectrum and network infrastructure to support the neutral host small cell model. The choices of spectrum and infrastructure are clearly intertwined. In this Section, we discuss the available and emerging alternatives. 

Spectrum resources could be provided by either the hosted operators or the SCPs. In case the spectrum used by the small cells is licensed to one or more of the hosted operators, the reference architecture is the Multiple Operators Core Network (MOCN) released by 3rd Generation Partnership Project (3GPP). 
%
While MOCN was originally released as part of Long-Term Evolution (LTE), 5G will rely on this architecture for network sharing \cite{3GPP2}.

The MOCN architecture envisages sharing in the radio access network only, where radio access nodes can connect to multiple core networks. Specifically, the evolved node B (eNodeB) broadcasts the Public land mobile network identifications (PLMN-IDs) of all the core networks it is connected to; the User Equipment (UE) signals to the eNodeB the PLMN-ID of its home network and the eNodeB routes the UE’s traffic to the corresponding core network. Changes adopted since Release 14 allow different operators to assign independent cell IDs and tracking area codes to shared radio nodes \cite{3GPP1}, thus removing the need for coordination among different core operators. Although the MOCN specifications were designed for macrocells sharing, this architecture is also considered a viable option for small cell sharing \cite{SF1}. Indeed, recently MOCN-based small cell solutions, e.g. SUMO \cite{ipAccess}, have appeared on the market. 


Since MOCN was designed for macrocell sharing, it will operate on spectrum licensed to one or more of the hosted operators. Licensed spectrum can be pooled from all/some of the hosted operators’ resources, or it can be used in an orthogonal manner - similar to the multi-operator radio access network solution (MORAN) - or a combination of orthogonal and shared resources can be implemented. In any case, an agreement between the neutral host provider and the hosted clients must be formed to allow the neutral host to access spectrum. 

Currently, there are at least three drawbacks with the use of licensed spectrum in small cell deployments provided by third parties. Firstly, under current licensing rules, due to competition concerns, in many countries spectrum sharing from different licensees is not permitted, or may only be allowed subject to regulators’ approval. Since competition concerns might be less relevant in the case of spectrum sharing for small cells, it is possible that appropriate regulation will be introduced to streamline the process for such agreements \cite{SF2}. Secondly, it is unlikely that a large MNO will invest in establishing agreements with neutral host providers that cover small areas with limited potential for return, if these arrangements involve access to licensed spectrum. Finally, using the operators licensed spectrum in the small cell network introduces an additional complicating factor, i.e. the coexistence between the macro and small tiers. 

The spectrum used by the small cells can also be provided by the SCPs. In this case, perhaps the most straightforward option is to use unlicensed spectrum. In addition to Hotspot 2.0 and Open SSID, MulteFire \cite{MulteFire} is another technology option in the 5 GHz unlicensed band. MulteFire is an LTE-based technology that operates in unlicensed spectrum without the support of licensed spectrum. The specifications, released in 2016, include a neutral host access mode whose reference architecture includes a neutral host core network. 

SCPs could also leverage recent developments in spectrum sharing frameworks to obtain access to spectrum resources. In particular, in 2015 the Federal Communications Commission (FCC) established the Citizens Broadband Radio Service (CBRS) for shared use of 150 MHz of spectrum in the 3.5 GHz band \cite{FCC1}. CBRS is based on a three-tiered sharing framework in which the highest tier, which includes federal and non-federal incumbent users, is protected from interference generated by the two lower tiers - Priority Access (PA) and General Authorized Access (GAA). PA users will be protected from interference generated by GAA users and other PA users, while GAA users will receive no interference protection. GAA use is the most straightforward way in which small neutral host providers can gain access to spectrum, albeit unprotected, within the CBRS framework. 
Alternatively, small neutral host providers could negotiate a leasing arrangement with one or more PA Licensees and acquire spectrum resources exclusively in a small geographic area . It should be noted that, in case the spectrum used by the small cells is obtained through GAA or PAL, MOCN-based small cell solutions are already available \cite{ipAccess}.

The ample range of options described above show that there is no dearth of opportunities in terms of access to spectrum to support the neutral host small cell model. Table~\ref{Table:1} summarises these spectrum opportunities.
\begin{table}[]
\centering

\begin{tabular}{l|l|l|ll}
\cline{2-3}
                                                                   & \textbf{Spectrum Access}                                                                 & \textbf{Spectrum provider} &  &  \\ \cline{1-3}
\multicolumn{1}{|l|}{\multirow{3}{*}{\textbf{Protected Access}}}   & \multirow{2}{*}{\begin{tabular}[c]{@{}l@{}}Nation-wide\\ exclusive license\end{tabular}} & MNOs (pooled)              &  &  \\ \cline{3-3}
\multicolumn{1}{|l|}{}                                             &                                                                                          & MNOs (orthogonal)          &  &  \\ \cline{2-3}
\multicolumn{1}{|l|}{}                                             & \begin{tabular}[c]{@{}l@{}}Local-area license\\ (e.g PAL)\end{tabular}                   & SCP or MNO                 &  &  \\ \cline{1-3}
\multicolumn{1}{|l|}{\multirow{2}{*}{\textbf{Unprotected Access}}} & \begin{tabular}[c]{@{}l@{}}Admission control\\ (e.g. GAA)\end{tabular}                   & SCP                        &  &  \\ \cline{2-3}
\multicolumn{1}{|l|}{}                                             & License Exempt                                                                           & --                         &  &  \\ \cline{1-3}
\end{tabular}\caption{Spectrum options for neutral host access}\label{Table:1}

\end{table}

\section{SMART CONTRACTS AS AN ALTERNATIVE TO SERVICE LEVEL AGREEMENTS (SLAS)}\label{Sec:SLA}
Regardless of the method chosen by the SCP to access spectrum to provide its services to the MNO, these two entities will need to enter a business agreement to regulate their interactions. 
%
%
Large businesses usually enter complex SLAs to define such agreements and the consequences of breaking them.
However, drafting an SLA is a potentially long and costly process, and thus an unattractive proposition for small-scale agreements like the ones we are considering in this paper. Blockchain technology and smart contracts  can then represent a quick and cost-effective alternative. Therefore, in the rest of this section we briefly introduce blockchain and smart contracts. 

\subsection{Blockchain}
In the most general case, blockchain is a distributed immutable append-only ledger. Blockchain uses cryptographic techniques to guarantee immutability and achieve consensus in the absence of trust between its users. 
Blockchains can be public, private or consortium (hybrid). Anyone can join a public blockchain and all the transactions are visible to all, while a private blockchain is owned by an organization and only authorized users can join and view transactions. In blockchain, a group of recent transactions are gathered in a block, and the block is added to the blockchain after it is approved by majority of the nodes over a procedure called consensus. The consensus mechanism is the heart of the blockchain since it defines its resilience against attacks and keeps its integrity. Depending on the consensus mechanism blockchains are classified into permissionless and permissioned. In the first group, any node can participate in the consensus;  the most popular consensus mechanisms in this group is proof of work, which is used by Bitcoin and Ethereum. In the permissioned blockchains, only a group of selected nodes participate in the consensus; Federated Byzantine Agreement is a popular model in permissioned blockchains. For more detail on blockchain and consensus mechanisms see \cite{IoTBlockchain18}. In our PoC, we use smart contracts in a permissionless public blockchain, Ethereum.
%

\subsection{Smart Contracts}
Originally proposed in 1996, smart contracts are experiencing a second life thanks to the advent of blockchains. The term “smart contract” itself does not have a universally accepted definition, but in the context of this work it refers to a block of code deployed on a blockchain platform, with the purpose of autonomously and transparently enforcing contractual clauses between agreeing parties \cite{clack2016smart}. A smart contract will typically have various functions that can be triggered in response to specific events, and whose invocation can result in the transfer of funds, the enforcement of penalties, etc. 
Smart contracts are supported in Ethereum blockchain platform and each node can deploy smart contracts with a specific contract account which can be addressed by other nodes. Deployed smart contracts can be the recipient of transactions and can initiate transactions. 
The main platform independent advantages and challenges of smart contracts are the following:

\subsubsection{Advantages} As the source code for these contracts is publicly visible, and the blockchain platform on which the code resides guarantees immutability, entities interacting through a smart contract can rest assured that no spurious behavior will be possible from any of the contracting parties, thus creating the foundation for trusted interactions in a trustless environment. Furthermore, smart contracts do not incur the costs associated to drafting a binding agreement through standard legal practices. Finally, the ability to quickly move funds through the cryptocurrency supporting the blockchain platform on which the smart contract resides can make these transactions cheaper to execute and faster to settle. 


\subsubsection{Challenges} 
%
The first issue is the legal validity of these agreements. As reported in \cite{murphy2016can}, whether smart contracts can give rise to legally binding contractual relations depends on several factors, including the nature of the smart contract (i.e. whether they include or operate in conjunction with contractual terms), the specific jurisdiction in which the contract applies etc. 
To mitigate this issue, it is advisable having dispute resolution mechanisms included in the smart contract code.
Furthermore, smart contracts, like all programs, are vulnerable to bugs and attacks. 
It is recommended to include fail-safe mechanisms in the contract, e.g. allowing the recovery of any outstanding balance in the event of the discovery of a vulnerability in the code.

%

Despite the mentioned challenges, the benefits highlighted above make for a very strong case for the adoption of smart contracts in all those scenarios where traditional legal agreements would be too costly or too cumbersome. 

\section{CASE STUDY: BLOCKCHAIN-ENABLED AGREEMENTS}\label{sec:PoC}
To demonstrate the feasibility of our proposed approach, we have implemented a prototype of an Ethereum-based 
, best-effort, pay-as-you-go neutral host smart contract infrastructure. We consider an SCP that operates a MOCN-based small cell network and provides radio access service to the users of an MNO. In this simple use-case, the neutral host SCP will be compensated proportionally to the amount of traffic served to the MNO’s customers. No minimum service levels are defined, although the MNO can always terminate a contract if the quality of the service provided by the SCP is inadequate. The rest of this section details the functioning of our platform; the code of the prototype is available online\footnote{https://bitbucket.org/edipascale/scaas\_react/}.

\begin{figure*}[t]
    \centering
    \includegraphics[width = 0.9\textwidth]{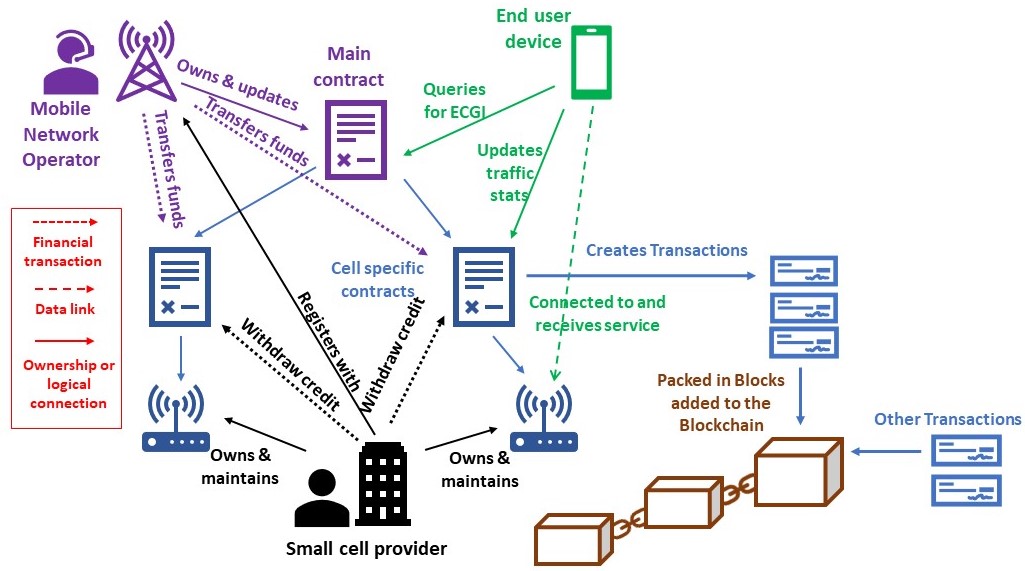}
    \caption{System architecture diagram for our smart contract infrastructure prototype.}
    \label{fig3}
\end{figure*}

    

As shown in Figure \ref{fig3}, the first step consists in the SCP registering as a provider with the MNO to receive the configuration parameters required to join its network. As part of the registration process, the SCP would also receive the global cell identifier (ECGI) to be broadcasted by the cell as part of the MNO’s infrastructure. In turn, the SCP would provide an Ethereum account address to be used to receive payments and interact with the contract. As a result, the “master” contract owned by the MNO will be updated with the association between the ECGI of the small cell and the contract address, so that it can be queried in the future by the end-user device. Furthermore, an instance of a contract between SCP and MNO for that specific cell will be spawned on the blockchain, capturing all the parameters of the agreement. In our PoC, this includes the cost per Kilobyte of traffic served by the small cell, the Ethereum address of the SCP, and the ECGI of the cell.

Each cell-specific instance of a contract will implement all the methods required for the billing and management of that particular agreement. Some of the methods will only be invokable by the MNO, e.g. to add funds to the contract which will be used to compensate the SCP. Other methods will only be callable by the SCP, e.g. to withdraw the credit accumulated for the served data. Ethereum allows contracts to fire events that can be intercepted by listening clients; this way, all the entities involved will be notified of any change to the state of the contract in near real time. 

Having the contract on the blockchain gives both the SCP and MNO the certainty that payment for the provided services will be carried out according to the agreements, thus adding transparency to the whole process. However, in the pay-as-you-go scenario we are considering, the smart contract has no built-in way of knowing the amount of traffic for which the SCP should be credited. This information could be provided by the MNO itself, but that would present a clear incentive to lie in the absence of trust. 


There are multiple ways in which this problem could be tackled; in our PoC, for example, we implemented an Android monitoring service running on the end-user device. Another way of tackling this problem is billing the MNO based on the number of UEs and their attachment time. Although this approach looks simpler, it creates higher number of transactions on the blockchain which is not desirable. 
The installation of the Android monitoring service could be incentivized by the MNO, e.g. as part of the software suite provided by many MNOs, or even be a requirement to make use of mobile data services in this augmented network. The purpose of this software is to act as a trusted oracle for the blockchain contract, providing accurate information on the data traffic serviced by the SCP on behalf of the MNO. Specifically, the service will query the main contract on the blockchain whenever the UE attaches to a new cell. If the ECGI of the cell is recorded in the contract, the latter will return the address of the cell-specific contract instance; the service will then track the data traffic consumed during this session, and update the contract when it moves to another cell. At this point the contract will credit the SCP for an amount proportional to the traffic served. The SCP, on the other hand, can verify the readings taken by the monitoring service to ensure that they match with its own measurements at the small cell. 


To prevent unauthorized traffic updates to the cell-specific contract, in our PoC the traffic-monitoring service comes with its own set of Ethereum credentials, which it uses to interact with the contract. As the method for updating the credit of the SCP is only accepting calls from this authorized set of credentials, the MNO can be assured that the information provided is truthful. In a real deployment, a more likely implementation would instead rely on user authentication (e.g. through one of the emerging digital identity services such as uPort) to track and verify the identity of the customer using the service. It should be noted that using blockchain/smart contracts will not add extra overhead to the capacity of the network as the blockchain is used for the billing system and not for  data communication. For example, the traffic update messages could be sent by UEs only when leaving a small cell.  Hence, the adoption of smart contracts would not affect the quality of experience of the network users.  However, using the proposed smart contract-based approach enables more SCPs to join and increases the opportunity of a UE being offloaded to a small cell with better received signal quality and lower load, as we show in the next section.

\begin{figure}
    \centering
    \includegraphics[width = .9\columnwidth]{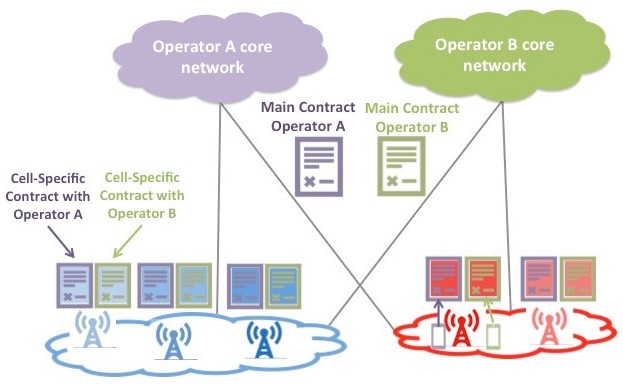}
    \caption{Multi-operator scenario: there is one master contract registered in the blockchain for each of the hosted operators. For each cell and for each MNO, a specific contract is spawned in the blockchain. The cell-specific contracts are depicted with two colors to highlight their association with the small cell and the MNO. }
    \label{fig4}
\end{figure}

The approach detailed above assumes the presence of a single MNO; from the perspective of the smart contract platform, multi-tenancy simply requires replicating these steps for each of the MNOs involved. Figure \ref{fig4} depicts the different components in the case of two SCPs (blue and red) hosting two MNOs (lilac and green).  

\begin{figure*}[t!]
    \centering
    \begin{subfigure}[t]{0.4\textwidth}
        \includegraphics[trim = 50 0 160 0 , clip, width=\textwidth]{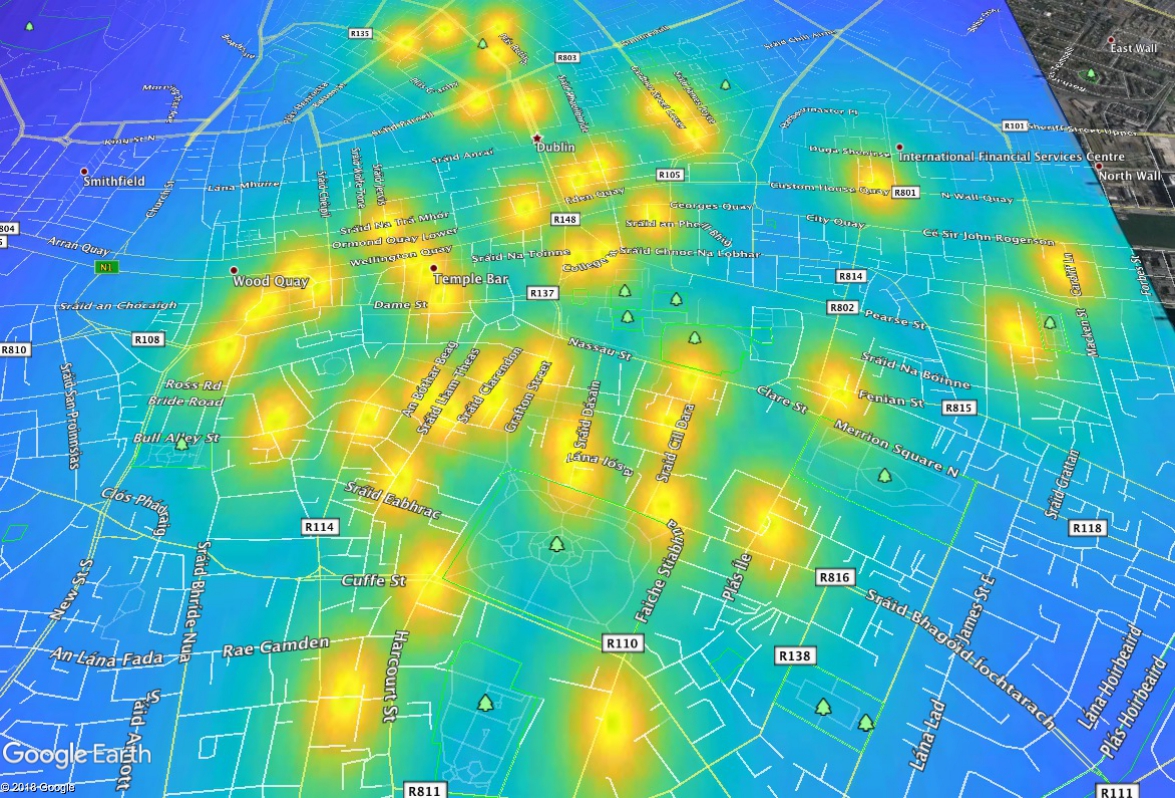}
        \label{fig:gull}
    \end{subfigure}
    \begin{subfigure}[t]{0.4\textwidth}
        \centering
        \includegraphics[trim = 50 0 160 0 , clip,width=\textwidth]{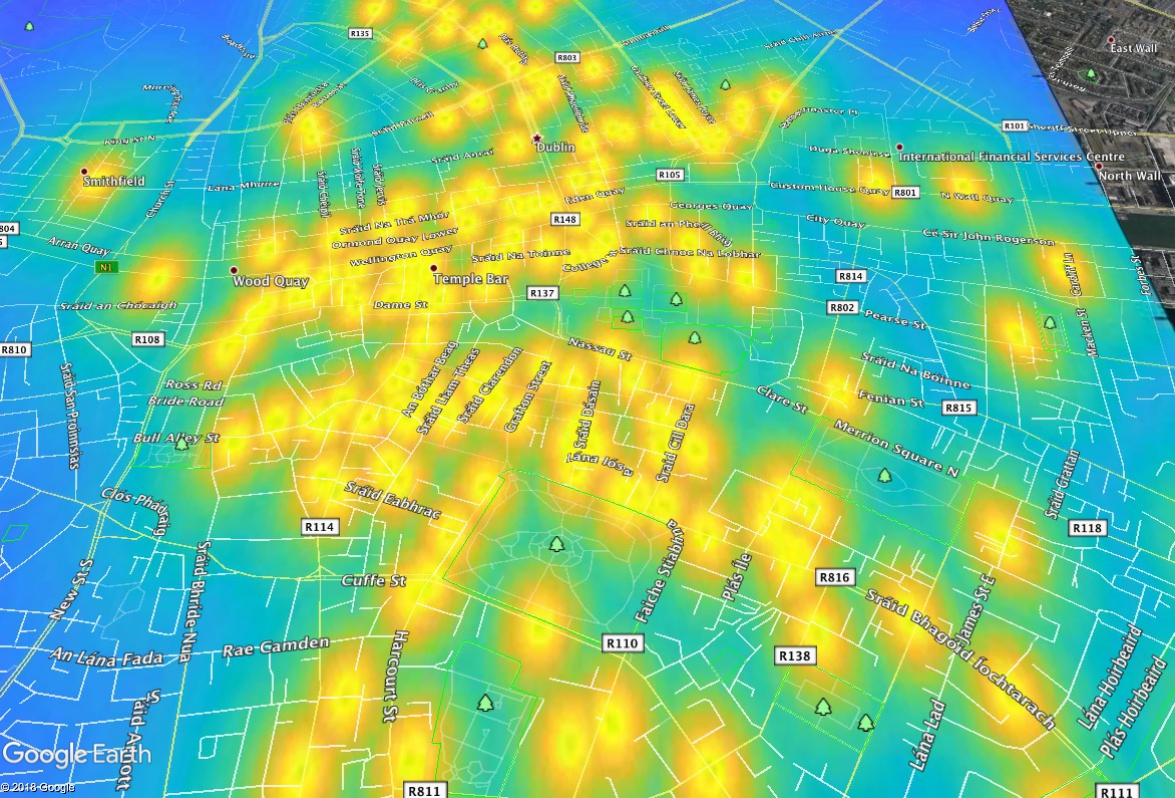}
    \end{subfigure}
    \begin{subfigure}[t]{0.15\textwidth}
        \centering
        \includegraphics[height = 170pt]{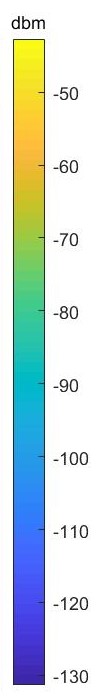}
    \end{subfigure}
    \caption{RSS in Dublin city centre from small cells deployed in all hotels (right) and chain hotels only (left). The list and locations of all hotels in the area was obtained using google place API. }
    \label{figure:Dublin}
\end{figure*}

\subsection{Formalization and enforcement of agreements between SCPs and MNOs}



While  our  case study is based on a best-effort
agreement between an SCP and an MNO, more sophisticated agreements could be implemented, e.g. based on E-UTRAN KPIs monitoring. 
However, the E-UTRAN KPIs only consider the E-UTRAN contribution: the core network normally relies on the counters provided by the cell, which, in a trustless environment, could be an unacceptable solution. An independent means of verification of the E-UTRAN KPIs via higher layer statistics that could be independently estimated by the MNO should be devised. The problem of how to ensure the synchronization of these measurements taken by different agents on different segments of the network, and how to reconcile the inevitable small differences that would emerge due to overheads, is an open question left for future work.

The smart contract that encodes the agreement between an MNO and an SCP could be extended to include clauses related to spectrum access. As pointed out by the FCC, the blockchain could be a key technology to enable spectrum sharing in the future. Indeed it could remove one of the impediments listed in Section II for the use of licensed spectrum in small cell deployments provided by third parties, namely the necessity of establishing legal arrangements regarding spectrum between the two parties. This could be done by: i) including spectrum access options and constraints in the smart contract that encodes the SLA between an MNO and an SCP; ii)  by creating a smart contract-based lightweight leasing agreement between an MNO and an SCP that could potentially interact with the SLA-based smart contract between the same parties.


With regards to penalties in case of a breach, there are many possible options: to name a few, a smart contract could implement monetary detractions (e.g. from the credit accumulated by the SCP), or automatic termination of the contract after a certain number of infractions. 

\subsection{Blockchain type and operational costs}
Due to Ethereum wide adoption and the simplicity of implementation, the described PoC is Ethereum-based, a public blockchain. It is important to point out that in a public chain the number of transactions-per-second is limited and transactions with low fees may stay pending for periods of time. 
Private/consortium blockchains, which can handle a much larger number of transactions-per-second using less computationally intensive consensus approaches, could be more suited to this particular application. It should be noted that the features implemented in the PoC are independent on the blockchain type.


Blockchain maintenance costs depend on the blockchain type. If public blockchains are used, each transaction has a small cost \cite{IoTBlockchain18}; if private/consortium blockchains are used, the costs depend on the required blockchain throughput, which is translated to processing power of the servers. The servers that host the blockchain can be maintained by the MNO or be rented from cloud service providers \cite{afraz2019distributed}. Creations of a smart contract will have a one-off cost of developing the master contract and making small modification for each specific one. Cancellation of a contract depends on the terms in the smart contract and it can be as simple as not offloading  UEs to the small cell or not accepting  new UEs at the small cell.  Renewal is as simple as addressing a revised smart contract.

\section{Implications for Network Service Improvement }\label{sec:implications}

We now investigate the benefits of  harnessing  small cell infrastructure with small to medium footprints.  
To present the potential that smart contracts can unlock, we analyzed the RSS improvement in Dublin city centre when hotels can operate as SCPs. 
We compared two scenarios. In scenario one, we assumed that in the absence of smart contracts agreements will be established only between MNOs and hotels that belong to chain holdings. In the second scenario, with smart contracts, all city centre hotels, including small hostels and B\&B, can operate as neutral host SCPs. Figure \ref{figure:Dublin} shows a comparison of the  received signal strength  for the two scenarios.   

\begin{figure}
    \centering
    \includegraphics[trim = 0 100 0 100 , clip,width = .9\columnwidth]{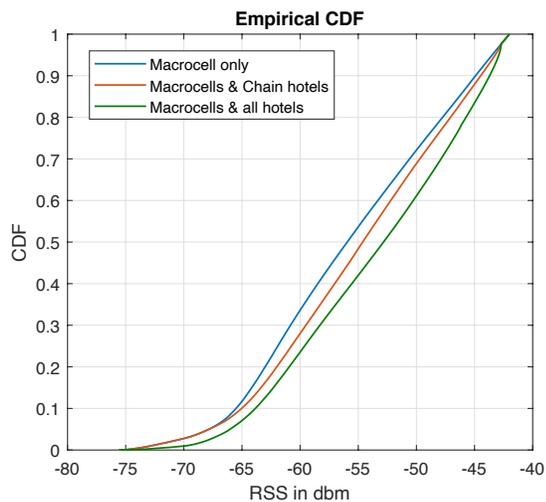}
    \caption{CDF of RSS where the $1\times1$ Kilometer area is covered by macrocells only, macrocells and small cells installed at chain hotels, and macrocells and small cells installed at all hotels.}
    \label{fig6}
\end{figure}

To quantify the impact on the network of an MNO, we computed the received signal map of the combined macro-small cell network with a resolution of $1\times1$ meter. In particular, we considered a square area of $1\times1$ Kilometer, south east corner at Latitude  $-6.2715$ and Longitude $53.337882$. For the macro tier, we used the BS locations of the operator with the largest number of BSs in the area.  For the small cell tier, we considered the same two scenarios presented in Figure \ref{figure:Dublin}.  Following 3GPP specifications, we set macro cell and small cell power to $46$ dbm and $24$ dbm, respectively. We considered pathloss and shadowing effects and averaged out fast fading. The RSS of each $1\times1$ meter location is the  strongest signal received at that location.
Since the RSS at the closest point to the small cell is $-42$ dbm the locations with RSS below this are our concern and Figure~\ref{fig6} compares the CDF of under $-42$ dbm RSS in the studied area, covering almost $64\%$ of the area. It highlights that being able to  incorporate in the MNO network all hotels in the area will result in an increase of the RSS of more than $10\%$ for almost all power levels. Our results show that the average increase in RSS is about $0.8$ dbm in scenario one, and it is around $2.3$ dbm in scenario two. Moreover, in scenario two the number of points with higher RSS compared to macro cell only scenario is three times larger than scenario one, showing another benefit of using smart contracts by enabling small and medium businesses to operate as neutral host SCPs.


\section{Conclusions}\label{Sec:conc}

In this paper, we proposed the adoptions of blockchain-enabled smart contracts to regulate agreements between MNOs and neutral host SCPs. The benefit of this approach is a significant reduction in time and cost with respect to the drafting of traditional SLAs. This, in turn, opens the market to  SCPs with small to medium footprints, and enables fine-grained agreements that otherwise would simply not be profitable. For MNOs the main incentive is to provide better coverage and Quality of Service while reducing operational and capital expenditure by partnering with SCPs using their neutral host small cells. For users, the incentive to install the traffic monitoring app is the ability to access an augmented network, which results in a higher Quality of Service.

To this aim, we have reviewed the numerous opportunities that exists for an SCP to access spectrum. Furthermore, to demonstrate the feasibility of the concept, a best-effort prototype, was implemented and tested on the public Ethereum Rinkeby test chain. With respect to the data traffic measurement, we showed that an app-based solution could be adopted, where the UEs provide information on the amount of traffic used to the smart contract. Since the measurements collected by the UEs – via an app distributed by the MNO – could be easily and independently verified by the SCP, a truthful interaction between the two parties is guaranteed. Finally, using the real network deployment and map data of city centre Dublin, Ireland, we have shown that the RSS will be improved on average by $10$ percent if neutral host small cells are deployed by all hotels. 


Future works will focus on extending the smart contract platform to support more elaborate levels of service. We will also investigate how market-based mechanisms can be used in the blockchain to dynamically determine the cost per Kilobyte of the traffic served by each small cell and how to encode spectrum access options and related constraints between MNOs and SCPs in a neutral host scenario.
\balance
\bibliographystyle{IEEEtran}
\bibliography{MyRef}
\section*{Biographies}
\begin{IEEEbiography}{Emanuele Di Pascale} received his PhD in Computer Science from Trinity College Dublin in 2015, for work on multimedia content delivery optimization over next-generation optical networks. He is currently working at Volta Networks. 
\end{IEEEbiography}
\begin{IEEEbiography}
{Hamed Ahmadi}
is an assistant professor in the department of Electronic Engineering at University of York, UK, and an adjunct assistant professor at University College Dublin, Ireland. He received his Ph.D. from National University of Singapore in 2012. Since then he worked at different academic and industrial positions in Ireland and UK. His research interests include design, analysis, and optimization of wireless communications networks, application of machine learning and Blockchain in wireless networks. 
\end{IEEEbiography}

\begin{IEEEbiography}{Linda Doyle} is the VP of Research /Dean of Research and Professor of Engineering and The Arts in Trinity College, University of Dublin. She was the founder Director of CONNECT a national research centre focused on future networks. Her expertise is in the fields of wireless communications, cognitive radio, spectrum management and creative arts practices. . She is a member of the National Broadband Steering Committee in Ireland, and is Chair of the Ofcom Spectrum Advisory Board in the UK. 
\end{IEEEbiography}

\begin{IEEEbiography}{Irene Macaluso} is a Senior Research Fellow at CONNECT, Ireland’s research centre for Future Networks and Communications, based at Trinity College, Dublin. She received her Ph.D. in Robotics from the University of Palermo in 2007. Her research focuses on the design of market-based mechanisms and the application of machine learning to radio resource sharing. She has published more than 70 papers in peer reviewed journals and conferences.
\end{IEEEbiography}

\end{document}